\documentclass[aps,twocolumn,amsmath,showpacs,amssymb,prb]{revtex4-1}
\usepackage{graphicx}
\usepackage{dcolumn}
\usepackage{amsmath}
\usepackage{color}
\usepackage{bm}

\begin{document}
\title{Transverse rectification in density-modulated two-dimensional electron gases}

\author{A.~Ganczarczyk}
\author{S.~Rojek}
\author{A.~Quindeau}
\author{M.~Geller}
\author{A.~Hucht}
\author{C.~Notthoff}
\affiliation{Faculty of Physics and CENIDE, University Duisburg-Essen, Lotharstra\ss e 1, 47057 Duisburg, Germany}
\author{D.~Reuter}
\author{A.~D.~Wieck}
\affiliation{Applied Solid State Physics, Ruhr-Universit\"at Bochum, Universit\"atsstra\ss e 150, 44780 Bochum, Germany}
\author{J.~K\"onig}
\author{A.~Lorke}
\affiliation{Faculty of Physics and CENIDE, University Duisburg-Essen, Lotharstra\ss e 1, 47057 Duisburg, Germany}

\date{\today}


\begin{abstract}
We demonstrate tunable transverse rectification in a density-modulated two-dimensional electron gas (2DEG). The density modulation is induced by two surface gates, running in parallel along a narrow stripe of 2DEG. A transverse voltage in the direction of the density modulation is observed, i.e. perpendicular to the applied source-drain voltage. The polarity of the transverse voltage is independent of the polarity of the source-drain voltage, demonstrating rectification in the device. We find that the transverse voltage $U_{y}$ depends quadratically on the applied source-drain voltage and non-monotonically on the density modulation. The experimental results are discussed in the framework of a diffusion thermopower model.
\end{abstract}

\maketitle
One of the most striking features of mesoscopic structures is the fact that their properties are no longer solely given by the underlying material properties, but also by their size and shape. As an example, the unipolar output of a ballistic rectifier \cite{Song1998} is not based on a change in material (such as a Schottky contact or the doping concentration in a $p$-$n$-junction) but on the introduction of a symmetry-breaking scatterer. Since this is a completely new concept for the realization of functional devices, quite some work has been devoted to understanding the basic principles that lead to rectification in symmetry-broken structures,\cite{Hieke2000, Shorubalko2001, Worschech2001, Hackens2004, Knop2006, Sassine2008} which have been shown to operate even at elevated temperatures\cite{Hieke2000, Song2001, Spanheimer2009} and high frequencies.\cite{Lorke1998, Song2001, Shorubalko2002, Spanheimer2009} While the original, intuitive model of ballistically guided electrons can already account for many qualitative features observed in the experiment,\cite{Song1998} further studies have shown that a number of different effects contribute to the observed rectification. These include quantum effects, predicted by Fleischmann \textit{et al.} \cite{Fleischmann2002} and observed by de Haan \textit{et al.},\cite{Haan2004} as well as thermoelectric contributions and self-gating.\cite{Salloch2009, Wiemann2010} Many of the investigated devices have quite elaborate designs in order to separate out a certain contribution or to optimize the rectification efficiency.

\begin{figure}[t]
\includegraphics[width=8cm]{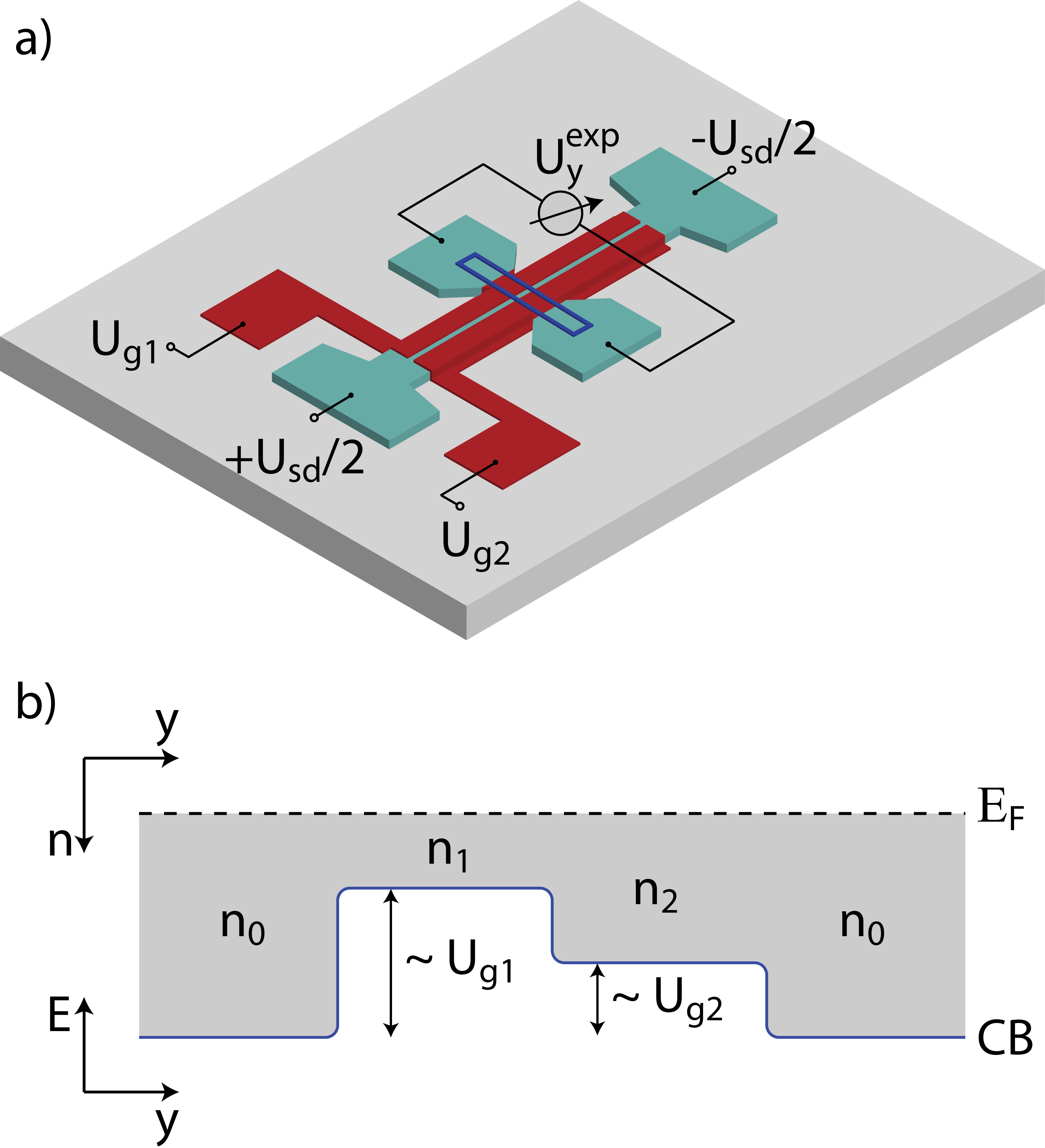}
\caption{\label{fig1}(a) Schematic picture of the rectifier and the experimental setup. The rectifier is measured in the push-pull configuration, i.e. half of the source-drain voltage $U_{\textrm{sd}}$ is applied at one terminal and the other half is applied at the second terminal. The source-drain voltages and the gate voltages $U_{\textrm{g}1}$ and $U_{\textrm{g}2}$ are both referenced to the ground potential. The transverse voltage $U_{y}^{\textrm{exp}}$ is measured between the left and right voltage probe. The 2DEG defined by an etched mesa is indicated in blue, the metallic gate electrodes in red. The carrier density in the area marked by the blue rectangle is sketched in (b) for $U_{\textrm{g}1} < U_{\textrm{g}2} < 0~\textrm{V}$, which gives a ratchet-type potential landscape. Here, $E_F$ marks the Fermi energy and CB the conduction band (dark-blue curve).}
\end{figure}

In the present work, we investigate transverse rectification in a two-dimensional electron gas (2DEG). By transverse, we mean that the unipolar voltage output is perpendicular to the applied input current. We choose a most simple design, which consists of two connected, parallel 2DEG stripes of different carrier density (see Fig.~\ref{fig1}(a)). Thus, the broken symmetry is only induced by the modulation of the carrier density and, contrary to earlier studies, the device is featureless along the current direction. In particular, no scatterers are induced, which guide or impede the flow of current. We find that the broken symmetry is sufficient to observe a clear rectification signal with a polarity that is given by the sign of the density difference. The experimental data are explained within a diffusion thermopower model that takes into account the spatial variation of the electron temperature and the thermopower.
The simplicity of the sample design allows for a direct comparison between theoretical results and experimental data. We find a good quantitative agreement without any adjustable parameters.

The samples are fabricated from a modulation-doped $\textrm{GaAs}$/$\textrm{AlGaAs}$ heterostructure grown by molecular beam epitaxy. Sample A is used for the rectification experiments, sample B is a reference sample. The samples contain a 2DEG, situated $107~\textrm{nm}$ below the surface. The charge carrier density $n_0$ and mobility $\mu_0$ at $T = 4.2~\textrm{K}$ are $2.07 \cdot 10^{15}~\textrm{m}^{-2}$ and $125~\textrm{m}^2/\textrm{Vs}$, respectively. Optical lithography and successive wet chemical etching is used to define a narrow stripe of 2DEG. Ohmic contacts are provided by evaporating Ni/AuGe/Au layers and successive thermal annealing. Two metallic gate electrodes, running parallel to the 2DEG stripe, are defined by electron beam lithography and consist of a $50~\textrm{nm}$ thick Au layer. A schematic picture of the resulting device (sample A) is shown in Fig.~\ref{fig1}(a). The 2DEG stripe has a length of $600~\mu \textrm{m}$ and a width of $50~\mu \textrm{m}$, while the length of the two metallic gate electrodes is $500~\mu \textrm{m}$ with a separation of $1~\mu \textrm{m}$. Different gate voltages $U_{\textrm{g}1}$ and $U_{\textrm{g}2}$ can be applied to the two gates, resulting in two stripes of different carrier density in the electron channel (see Fig.~\ref{fig1}(b)). The gap region between the gates equals the width of the transition region between areas of different carrier densities and will therefore be neglected in the following. The current is applied between the source and the drain contacts at the ends of the 2DEG stripe. The  transverse voltage $U_{y}^{\textrm{exp}}$  is measured by two voltage probes located on each side of the stripe. The measurements are performed in the push-pull configuration,\cite{Knop2006} i.e. half of the source-drain voltage $U_{\textrm{sd}}$ is applied at the source and the other half at the drain terminal. The outcome is a symmetric configuration with respect to the voltage probes. The reference sample B is used to determine the relevant sample parameters for the diffusion thermopower model through resistivity, Hall, and Shubnikov-de Haas measurements. In contrast to sample A, sample B has only one gate covering the whole stripe and three voltage probes on each side. If not stated otherwise, all measurements are performed on sample A at $4.2~\textrm{K}$.

\begin{figure}
\includegraphics[width=8cm]{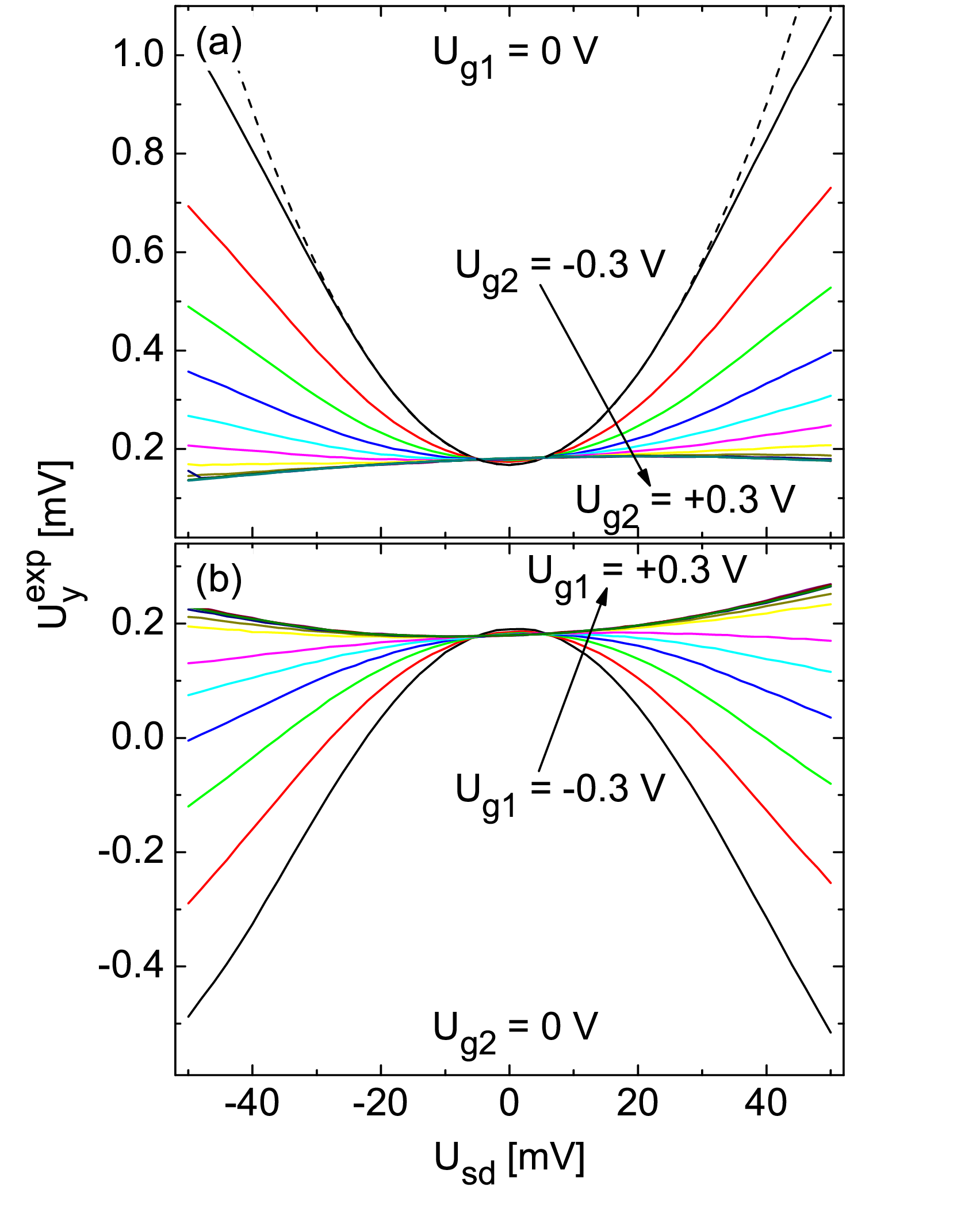}
\caption{\label{fig2}(a) Dependence of the transverse voltage $U_{y}^{\textrm{exp}}$ on the source-drain voltage $U_{\textrm{sd}}$. The gate voltage $U_{\textrm{g}2}$ is applied from $-0.3~\textrm{V}$ (black trace) to $+0.3~\textrm{V}$ (dark green trace) in steps of $0.05~\textrm{V}$, while gate 1 remains grounded ($U_{\textrm{g}1} = 0~\textrm{V}$). The yellow trace shows the transverse voltage for $U_{\textrm{g}2} = 0~\textrm{V}$. The dashed, black line is an exemplary parabolic fit for $U_{\textrm{sd}} \leq 25~\textrm{mV}$. (b) Transverse voltage $U_{y}^{\textrm{exp}}$, when the role of the gates is switched, i.e. the gate voltage $U_{\textrm{g}1}$ is varied, while gate 2 remains grounded.}
\end{figure}

Figure \ref{fig2}(a) shows the transverse voltage $U_{y}^{\textrm{exp}}$ as a function of the source-drain voltage $U_{\textrm{sd}}$. The gate voltage $U_{\textrm{g}2}$ is varied from $-0.3~\textrm{V}$ to $0.3~\textrm{V}$ in steps of $0.05~\textrm{V}$, while the second gate remains grounded ($U_{\textrm{g}1} = 0~\textrm{V}$). The transverse voltage exhibits an almost parabolic dependence on the source-drain voltage, demonstrating transverse rectification in the device. For a gate voltage of $U_{\textrm{g}2}=-0.3~\textrm{V}$, e.g., we find transverse voltages $U_{y}^{\textrm{exp}} = 1.04~\textrm{mV}$  and  $1.07~\textrm{mV}$ for $U_{\textrm{sd}}=-50~\textrm{mV}$ and $50~\textrm{mV}$, respectively. We attribute this negligible deviation from the symmetry relation $U_{y}^{\textrm{exp}}(U_{\textrm{sd}}) = U_{y}^{\textrm{exp}}(-U_{\textrm{sd}})$ to small imperfections of the device geometry such as slightly shifted voltage probes and a minute misalignment angle between the gates and the etched channel. Figure \ref{fig2}(a) also shows that the transverse voltage exhibits a strong dependence on the gate voltage $U_{\textrm{g}2}$, which will be discussed later in detail. In Fig.~\ref{fig2}(b) the role of the gates is switched, i.e. $U_{\textrm{g}1}$ is varied, while $U_{\textrm{g}2}$ remains constant. The polarity of the transverse voltage $U_{y}^{\textrm{exp}}$ has changed, since the direction of the density difference in the electron channel is reversed. In both cases, the polarity of the transverse voltage $U_{y}^{\textrm{exp}}$ corresponds to an induced flow of electrons from the high-carrier-density area to the low-density area.

\begin{figure}
\includegraphics[width=8cm]{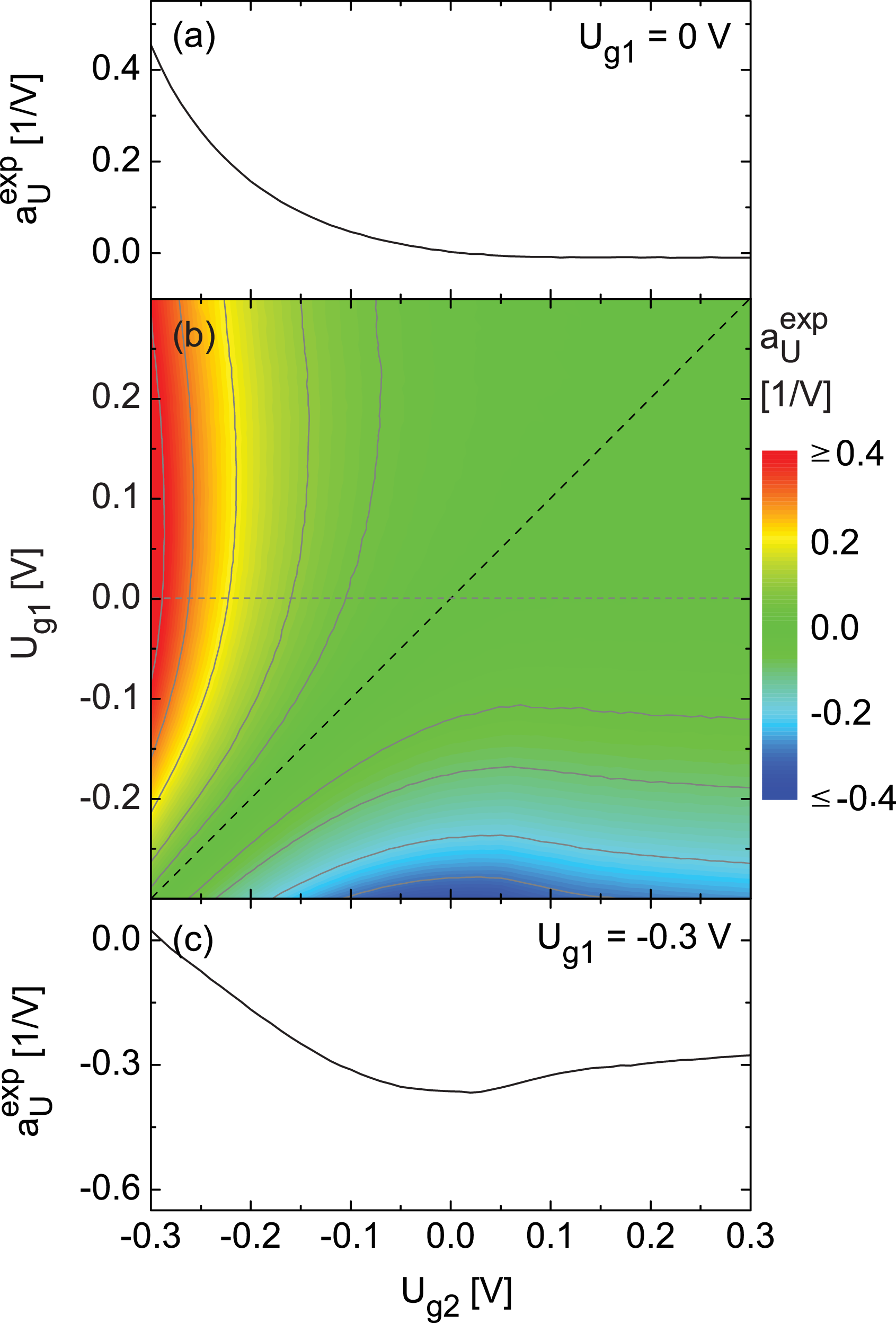}
\caption{\label{fig3}(a) Rectification efficiency $a_U^{\textrm{exp}}$ as a function of $U_{\textrm{g}2}$ for $U_{\textrm{g}1} = 0~\textrm{V}$, where $a_U^{\textrm{exp}}$ is the coefficient of a $a_U^{\textrm{exp}} \cdot {U^2_{\textrm{sd}}}$-fit on the traces shown in Fig.~\ref{fig2} for $U_{\textrm{sd}}$ smaller than $25~\textrm{mV}$. (b) Contour plot of the rectification efficiency $a_U^{\textrm{exp}}$ as a function of $U_{\textrm{g}1}$ and $U_{\textrm{g}2}$. The black dashed line represents $U_{\textrm{g}1} = U_{\textrm{g}2}$ and the gray dashed line shows the cross section, which is plotted in (a). The distance between the contour lines is $0.10~\textrm{V}^{-1}$ with additional contour lines at $\pm 0.05~\textrm{V}^{-1}$. (c) $a_U^{\textrm{exp}}$ in dependence on $U_{\textrm{g}2}$ for $U_{\textrm{g}1} = -0.3~\textrm{V}$. This trace matches the cross section of the contour plot in (b) at $U_{\textrm{g}1} = -0.3~\textrm{V}$.}
\end{figure}

To better quantify how the rectification depends on different parameters, we introduce the rectification efficiency. It is defined as the pre-factor $a_U^{\textrm{exp}}$ in a parabolic fit to the experimental data $U_{y}^{\textrm{exp}}=a_U^{\textrm{exp}} \cdot {U_{\textrm{sd}}^2}$ for a bias range $U_{\textrm{sd}} \leq 25~\textrm{mV}$ (dashed line in Fig.~\ref{fig2}). The rectification efficiency is determined for a wide range of gate voltage combinations and the results are presented in Fig.~\ref{fig3}.

Figure \ref{fig3}(a) shows $a_U^{\textrm{exp}}$ as a function of $U_{\textrm{g}2}$ for $U_{\textrm{g}1} = 0~\textrm{V}$. As already observed in Fig.~\ref{fig2}, maximum rectification efficiency is seen around $U_{\textrm{g}2} = -0.3~\textrm{V}$, i.e. close to the depletion voltage of the sample. With increasing $U_{\textrm{g}2}$, $a_U^{\textrm{exp}}$ decreases strongly at first and switches polarity at approx.~$U_{\textrm{g}2} = 0~\textrm{V}$. For small positive gate voltages, $|a_U^{\textrm{exp}}|$ increases slowly and saturates for high positive gate voltages ($U_{\textrm{g}2} > 0.1~\textrm{V}$).

The contour plot in Fig.~\ref{fig3}(b) gives an overview over $a_U^{\textrm{exp}}$ in the complete range of investigated gate voltages $U_{\textrm{g}1}$ and $U_{\textrm{g}2}$. It can be seen that the rectification efficiency is antisymmetric under exchanging $U_{\textrm{g}1}$ and $U_{\textrm{g}2}$, and, therefore, vanishes for $U_{\textrm{g}1} = U_{\textrm{g}2}$ (black dashed line). The maximum of $|a_U^{\textrm{exp}}|$ ($\approx 0.4~\textrm{V}^{-1}$) is observed at roughly $U_{\textrm{g}1} = 0~\textrm{V}$ and $U_{\textrm{g}2} = -0.3~\textrm{V}$ or vice versa. This can be more clearly seen in Figs.~\ref{fig3}(a) and \ref{fig3}(c), which show sections through the data along the $U_{\textrm{g}1} = 0~\textrm{V}$ and the $U_{\textrm{g}1} = -0.3~\textrm{V}$ line, respectively. The small difference between $|a_U^{\textrm{exp}} (0~\textrm{V},-0.3~\textrm{V})|$ and $|a_U^{\textrm{exp}}(-0.3~\textrm{V}, 0~\textrm{V})|$ is again attributed to imperfections in the sample processing.

The rectification efficiency in Fig.~\ref{fig3}(c) exhibits a non-monotonic dependence on the gate voltage $U_{\textrm{g}2}$. This is unexpected at first, because the carrier density difference monotonically increases with $U_{\textrm{g}2}$. The fact that $|a_U^{\textrm{exp}}|$ \emph{decreases} while the density step height increases in the bias range $U_{\textrm{g}2} > 0$ is an indication that a purely ballistic picture cannot explain the observed rectification. The ballistic deflection of electrons in density modulated systems \cite{Spector1990, Spector1990a, Noguchi1993, Fukai1992} increases monotonically with increasing step height. So, even though a simple ballistic model may be able to account for some qualitative features of our experiment,\cite{Ganczarczyk2009a} a more in-depth treatment is necessary.

\begin{figure}
 \includegraphics[width=8cm]{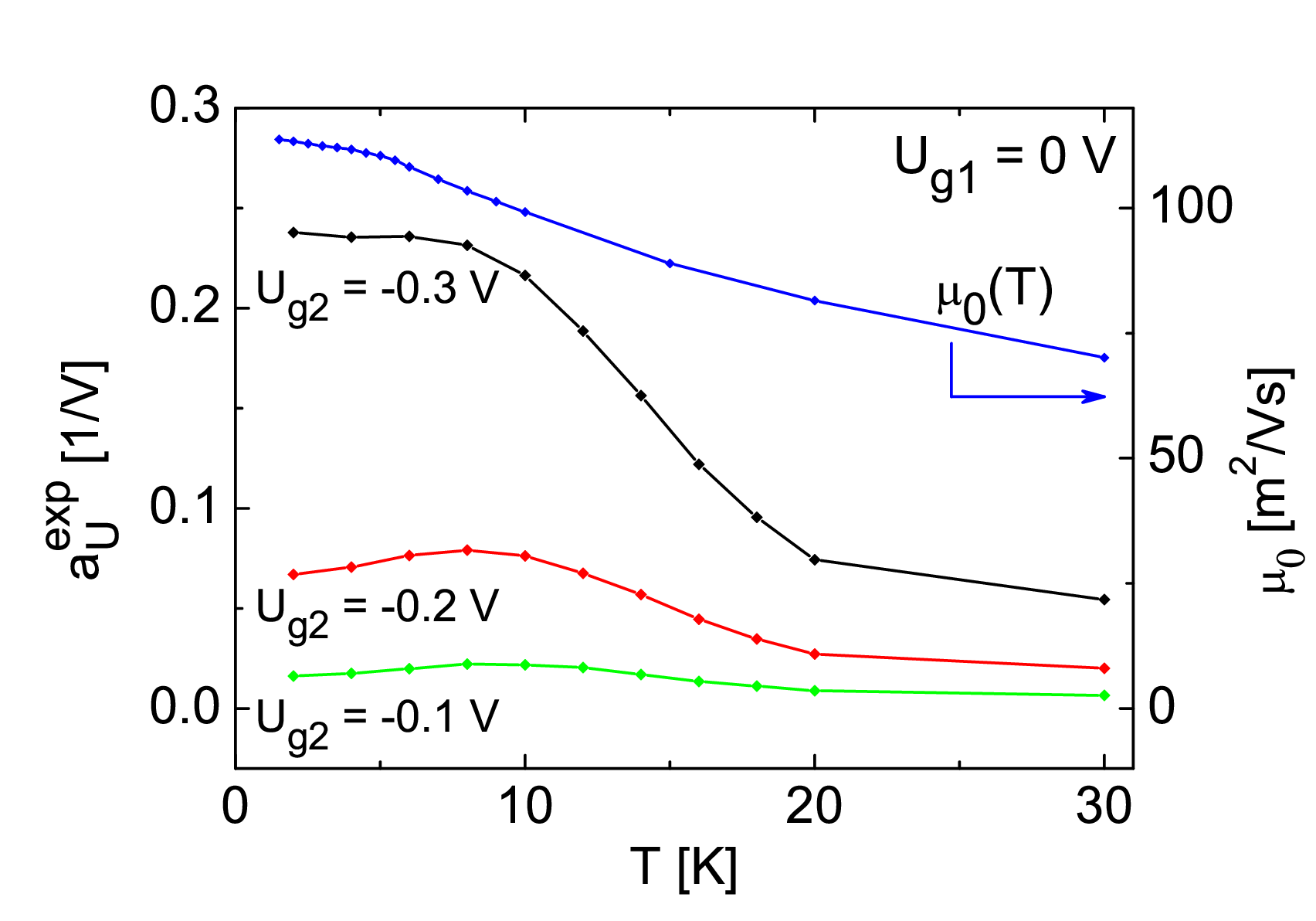}
 \caption{\label{fig4}Temperature dependence of the rectification efficiency $a_U^{\textrm{exp}}$ (left axis). The black, red and green traces show $a_U^{\textrm{exp}}$ as a function of the temperature $T$ for $U_{\textrm{g}2} = -0.3~\textrm{V}$, $U_{\textrm{g}2} = -0.2~\textrm{V}$ and $U_{\textrm{g}2} = -0.1~\textrm{V}$, respectively. The blue trace shows the temperature dependence of the mobility $\mu$ measured on the reference sample B (right axis).
 }
\end{figure}

Additional insight into the rectification mechanism comes from its temperature dependence. In  Fig.~\ref{fig4} the temperature dependence of the rectification efficiency $a_U^{\textrm{exp}}$ (left axis) and  carrier mobility $\mu_0$  (right axis, measured on sample B) are presented. The black, red and green traces show $a_U^{\textrm{exp}}$ in the range from $2~\textrm{K}$ to $30~\textrm{K}$ for the gate voltages $U_{\textrm{g}2} = -0.3~\textrm{V}$, $U_{\textrm{g}2} = -0.2~\textrm{V}$ and $U_{\textrm{g}2} = -0.1~\textrm{V}$, respectively, while $U_{\textrm{g}1} = 0~\textrm{V}$. The blue trace shows the temperature dependence of the mobility $\mu$, which was determined using Hall- and resistivity measurements. While the mobility decreases roughly linearly, the rectification efficiency shows a more complex behavior, in particular for $U_{\textrm{g}2} = -0.2~\textrm{V}$ and $U_{\textrm{g}2} = -0.1~\textrm{V}$, where $a_U^{\textrm{exp}}$ exhibits a maximum at approx. $8-10~\textrm{K}$. This again indicates that the observed rectification is not a purely ballistic effect, which would directly reflect the electron mobility or, equivalently, the elastic mean free path.

\begin{figure}
 \includegraphics[width=8cm]{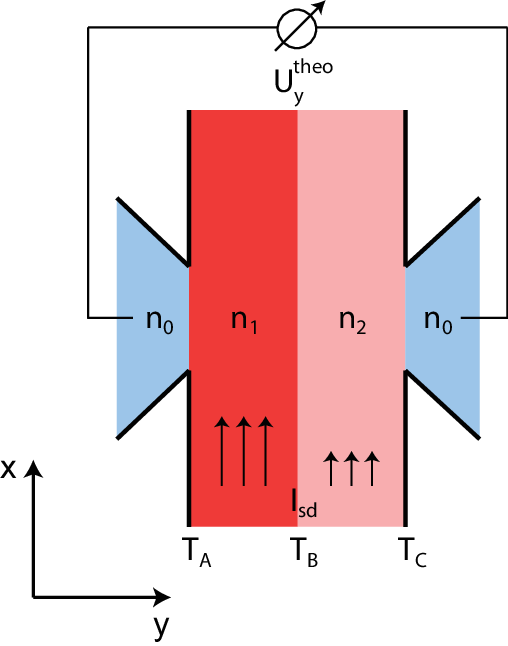}
 \caption{\label{fig5}Schematic sketch of the rectifier. The system is divided into four regions: The two stripes of the electron channel plus the two voltage probes. The current-free voltage probes have the carrier density $n_0$, the two stripes have (tunable) carrier concentrations $n_1$ and $n_2$.}
\end{figure}

In the following, we show that the measured transverse voltage $U^{\textrm{exp}}_{y}$ can be explained within a diffusion thermopower model.
The relevant part of the rectifier is schematically shown in Fig.~\ref{fig5}. We distinguish different regions: the electron channel, through which a finite current is driven, consists of two stripes 1 and 2 with, in general, different carrier concentrations $n_1$ and $n_2$. The voltage probes are current free and remain at carrier concentration $n_0$, independent of the gate voltages. The current along the electron channel increases the local electron temperature from the lattice temperature $T_0$ to $T=T_0 + \delta T$ via Joule heating. The density modulation in the perpendicular direction of the electron channel introduces both a $y$-dependence of the temperature $T(y)$ (via the carrier-density dependence of the resistivity) and of the Seebeck coefficient $S(y)$, which characterizes the strength of the thermopower. This results in a finite thermoelectric voltage
\begin{align}
	U^{\textrm{theo}}_{y}= \int^\infty_{-\infty} \text{d}y\ S(y)\  \partial_y T(y)\; , \label{MottInt}
\end{align}
between the contacts of the voltage probes that are far outside the heated region, $T(\infty) = T(-\infty) = T_0$. Neglecting the $y$-dependence of the Seebeck coefficient in Eq.~(\ref{MottInt}) would lead to an exact cancellation of the contributions from regions with positive and negative temperature gradients. However, when taking the $y$-dependence of $S$ into account, the region with the smaller carrier concentration dominates (since it has a larger Seebeck coefficient), and the thermoelectric voltage, Eq.~(\ref{MottInt}), is non-vanishing.

For a quantitative analysis, we employ a semiclassical transport theory.\cite{Ashcroft1976}
The $y$-components of the electric and the heat current densities, $j_y$ and $j^q_y$, respectively, are related to the gradients of the electrochemical potential $\phi$ and temperature $T$ via
\begin{equation}
	\left( \begin{array}{c} j_y \\ j^q_y \end{array} \right)
	=
	- \left( \begin{array}{cc} L_{11} & L_{12} \\ 
	L_{21} & L_{22}  \end{array} \right)
	\left( \begin{array}{c} \partial_y \phi \\ \partial_y T \end{array} \right)
	\, .
\end{equation}
The coefficients $L_{ij}$ are related to the electric conductivity $\sigma$, the Seebeck coefficient $S$ and the thermal conductivity $\kappa$ via $L_{11}=\sigma$, $S=L_{12}/L_{11}$, $L_{21}=TL_{12}$, and $\kappa=L_{22}-L_{12}L_{21}/L_{11}$.

The electron temperature profile is determined by the continuity equation for the heat current,
\begin{align}
	\nabla \cdot {\mathbf{j}^q} = \mathbf{j} \cdot \mathbf{E}-c_V \frac{T-T_0}{\tau_\textrm{i}} \, .
\label{continuity}
\end{align}
The first (source) and second (drain) term on the right hand side account for Joule heating and heat transfer to the lattice, respectively. The latter is modeled by a relaxation term that includes the specific heat capacity $c_V$, the difference between the electron ($T$) and lattice ($T_0$) temperature, and a heat relaxation time $\tau_\textrm{i}(\epsilon)$ taken at the Fermi energy, $\tau_\textrm{i}=\tau_\textrm{i}(\epsilon_F)$. We remark here, that the relaxation time $\tau_\textrm{i}$ for the heat transfer to the lattice differs from the relaxation time $\tau$ entering the electric conductivity. For $\tau_\textrm{i}$, only those electron scattering processes are relevant which are accompanied with an energy change larger than $k_BT$ (dubbed as {\em inelastic} scattering in contrast to {\em elastic} scattering for which the energy transfer is much smaller than $k_BT$).\cite{Kawamura1992} In contrast, both elastic and inelastic scattering processes contribute to the (energy-dependent\cite{Walukiewicz1984, Harris1989}) relaxation time entering the electric conductivity, $\tau^{-1}(\epsilon)=\tau^{-1}_\textrm{e}(\epsilon) + \tau^{-1}_\textrm{i}(\epsilon)$. In the system described here, elastic scattering strongly dominates, $\tau_\textrm{e} \ll \tau_\textrm{i}$, and we can set $\tau\approx \tau_\textrm{e}$.

Since the charge current density has only an $x$-component, which does not depend on the position $x$, we simplify Eq.~(\ref{continuity}) by using $\nabla \cdot {\mathbf{j}^q} = \partial_y j_y$ and $\mathbf{j} \cdot \mathbf{E} = \sigma E_x^2$, combine it with the definition of the thermal conductivity, $j_y^q= -\kappa\; \partial_y T$, to obtain a differential equation  
\begin{align}
	T-l^2 \partial^2_y T = \frac{(e E_x l)^2}{\kappa/\sigma} + T_0 \, ,
	\label{Tdiff}
\end{align}
that describes the variation of the temperature on the scale given by the energy diffusion length $l=\sqrt{\kappa \tau_\textrm{i}/c_V}$, which is considerably larger than the mean free path in the 2DEG.

Calculating $\sigma$, $\kappa$, and $c_V$ to lowest order in the Sommerfeld expansion (valid for $k_BT \ll \epsilon_F$) yields the energy diffusion length
\begin{align}\label{energydiffusion}
	l = v_F\sqrt{\frac{\tau_\textrm{i} \tau_\textrm{e}}{2}} \, ,
\end{align}
with $\tau_\textrm{e} = \tau_\textrm{e}(\epsilon_F)$, and the Wiedemann-Franz law $\kappa/\sigma= (\pi k_B/e)^2T/3$. We use $v_F= \sqrt{2\epsilon_F/m^\ast}$, where $m^\ast$ is the effective electron mass in GaAs. Furthermore, the Fermi energy $\epsilon_F$ (relative to the lower subband edge) is related to the two-dimensional density $n$ of the electron gas via $\epsilon_F = n\pi \hbar^2/m^\ast$. This expression also relates the carrier density to the gate voltages $U_{\textrm{g}1}$ and $U_{\textrm{g}2}$ via $\epsilon_F/\epsilon_0=1+U_{\textrm{g}i}/U^\textrm{dep}_{\textrm{g}}$, where $\epsilon_0$ is the Fermi energy at zero gate voltage and $U^\textrm{dep}_{\textrm{g}} = -0.38~\textrm{V}$ the depletion voltage.

Applying the Sommerfeld expansion, the Seebeck coefficient, $S$, can be expressed in terms of the electric conductivity by the Mott formula\cite{Cutler1969, Ashcroft1976}
\begin{align}
 	S =-\frac{\pi^2}{3}\frac{k_B^2T}{e}\frac{\sigma^\prime}{\sigma},
 \label{thermopower}
\end{align} 
with $\sigma = \sigma(\epsilon_F)$, $\sigma^\prime = \left.\partial_\epsilon \sigma(\epsilon)\right|_{\epsilon_F}$, where (for a two-dimensional electron gas) $\sigma(\epsilon) = e^2 \epsilon \tau_\textrm{e}(\epsilon)/(\pi \hbar^2)$.

For a systematic perturbation expansion up to second order in $E_x$, we replace $T$ by $T_0$ in the ratio $\kappa/\sigma$ and in the Mott formula, Eq.~\eqref{thermopower}. Then, the solution of Eq.~(\ref{Tdiff}) is given by 
\begin{equation}
	T_i(y) =T^{\textrm{bulk}}_i + T^-_i e^{-y/l_i} + T^+_i e^{y/l_i}
\end{equation}
for each region $i$ of constant carrier density. The asymptotic bulk value $T^\textrm{bulk}_i$ is reached far away from the steps: $T^\textrm{bulk}_0 = T_0$ deep in the voltage probes and $T^\textrm{bulk}_i = T_0 + (e E_x l_i)^2/(\frac{\pi^2}{3}k_B^2T_0)$ with the corresponding energy diffusion length $l_i$ for the two stripes $i = 1$ and $i = 2$ in the electron channel, respectively.

The constants $T^\pm_i$ for each stripe are determined by the boundary conditions. To fully determine all constants, we need two conditions for each interface between regions of different carrier concentration. One is readily obtained by requiring that the heat current $j_y^q = - \kappa\partial_y T$ entering an interface from one side equals the one leaving to the other side, which for the system under study yields that the product $n \tau_\textrm{e} \partial_y T$ is continuous at each interface. As a second condition we find that, for $\tau_\textrm{i} \gg \tau_\textrm{e}$, the temperature itself is continuous. \footnote{This is proven by writing the electric and the heat current through each interface as functions of the steps in temperature, $\Delta T$, and electrochemical potential,  $\Delta \phi$, and linearizing in $\Delta T$ and $\Delta \phi$. The resulting expressions depend on the transmission properties at the interface but are independent of $\tau_\textrm{i}$ and $\tau_\textrm{e}$. The requirement of a vanishing electric current relates $\Delta T$ to $\Delta \phi$ in a linear way, i.e., also the heat current is a linear function of $\Delta T$. This heat current has to match the heat current $j_y^q=-\kappa \partial_y T$ entering the interface, which scales with the amplitude of the temperature increase $\delta T$ times $\kappa/l \propto \sqrt{\tau_\textrm{e}/\tau_\textrm{i}}$. Therefore, $\Delta T$ scales with $\sqrt{\tau_\textrm{e}/\tau_\textrm{i}}$ times the amplitude of $\delta T$ and can, as a consequence, be neglected in the limit $\tau_\textrm{i} \gg \tau_\textrm{e}$. This also justifies Eq.~(\ref{thermopower}), which determines the thermopower everywhere away from the interfaces only, with no extra contributions from the interfaces.}

In order to calculate the thermoelectric voltage, it is necessary to determine the (energy-dependent) relaxation rates $\tau_\textrm{e}(\epsilon)$ and $\tau_\textrm{i}(\epsilon)$. A theoretical estimate of these relaxation rates would depend on various model assumptions and, thus, introduce uncertainties that make a quantitative comparison with experiment difficult. Therefore, we determine them experimentally by performing (gate-voltage dependent) resistivity, Hall, and Shubnikov-de Haas measurements on reference sample B.

The inelastic relaxation time $\tau_\textrm{i}$ is determined from the relation between electron temperature $T$ and source-drain current $I_{\textrm{sd}}$ in the heat balance equation Eq.~\eqref{continuity}, after we set $\nabla \cdot {\mathbf{j}^q}=0$.\cite{Molenkamp1990a} The latter is justified since, for all $x$-positions at which there is no voltage probe, the temperature and, thus, the heat current is independent of $y$ for sample B. To measure $T(I_{\textrm{sd}})$, we use the temperature dependence of the Shubnikov-de Haas (SdH) oscillations of two-dimensional electrons under low magnetic fields.\cite{Klitzing1980, Ando1982, Hirakawa1986a} This is done by first measuring the SdH oscillations at different lattice temperatures $T_0$ with a small current $I_{\textrm{sd}}$ of $100~\textrm{nA}$, such that the electron temperature $T$ equals the lattice temperature. Then, the SdH oscillations are measured at a fixed lattice temperature $T_0$ with different heating currents $I_{\textrm{sd}}$. The comparison of the SdH-oscillation amplitudes in the two different schemes yields $T(I_{\textrm{sd}})$. An alternative approach to determine $T(I_{\textrm{sd}})$ would be to measure the temperature-dependent resistivity.\cite{Manion1987} Using the SdH oscillations, however, has the advantage that its amplitude depends only on the electron but not the lattice temperature, in contrast to the resistivity that, in general, depends on both the electron and the lattice temperature.

\begin{figure}
 \includegraphics[width=8cm]{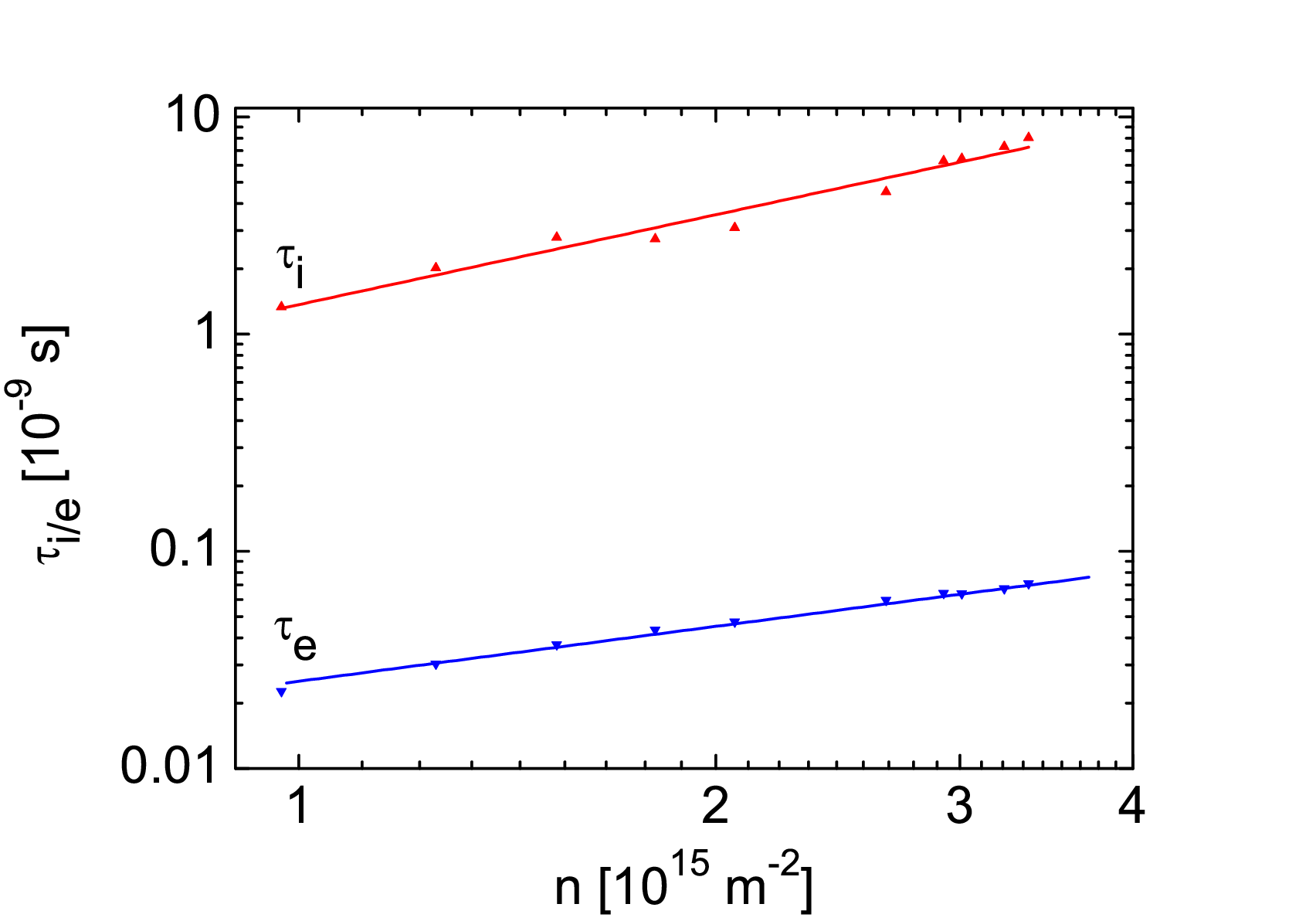}
 \caption{\label{fig6}Energy dependence of the elastic scattering time $\tau_\textrm{e}$ (blue) and the inelastic scattering time $\tau_\textrm{i}$ (red). The first data points match $U_{\textrm{g}2} = -0.2~\textrm{V}$ and the gate voltage increases by $0.05~\textrm{V}$ for each data point. The lines correspond to the linear fits of the data.}
\end{figure}

The results for $\tau_\textrm{e}(\epsilon)$ and $\tau_\textrm{i}(\epsilon)$ are shown in Fig.~\ref{fig6}. Fitting a power-law function, $\tau_\textrm{e} = \tau_{\textrm{e},0} \left(\epsilon/\epsilon_0\right)^{\alpha_\textrm{e}}$ and $\tau_\textrm{i} = \tau_{\textrm{i},0} \left(\epsilon/\epsilon_0\right)^{\alpha_\textrm{i}}$ yields $\tau_{\textrm{e},0}=4.7 \cdot 10^{-11}~\textrm{s}$ and $\tau_{\textrm{i},0}= 3.7 \cdot 10^{-9}~\textrm{s}$, and for the exponents, $\alpha_\textrm{e} = 0.88 \pm 0.04$ and $\alpha_\textrm{i} = 1.45 \pm 0.09$, in good agreement with theoretical predictions.\cite{Hwang2008a}

\begin{figure}
 \includegraphics[width=8cm]{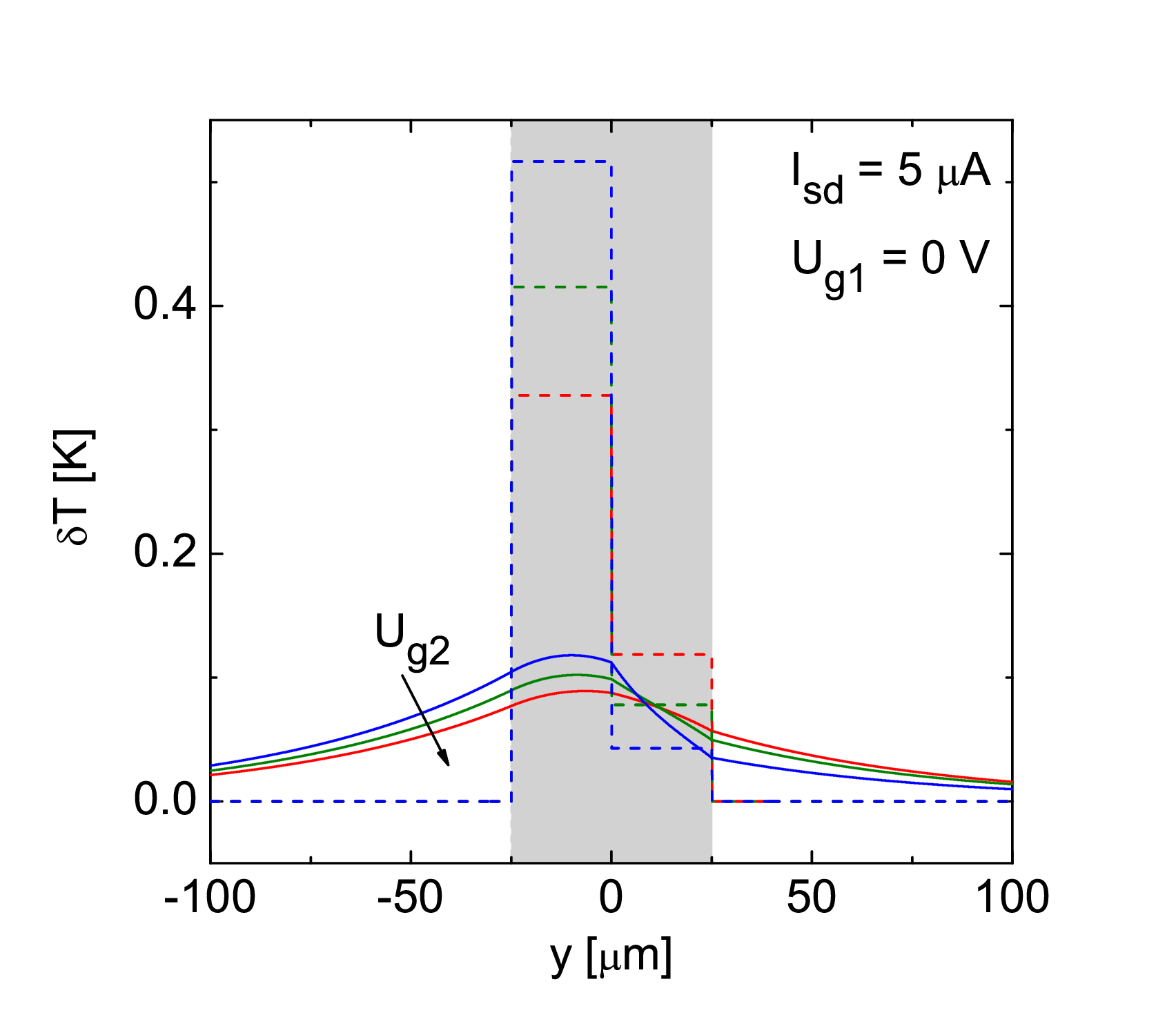}
 \caption{\label{fig7}Heating correction to the temperature of the electron system for $I_{\textrm{sd}}= 5~ \mu\textrm{A}$, $U_{\textrm{g}1} = 0~\textrm{V}$, and three different values for $U_{\textrm{g}2}$: $U_{\textrm{g}2}=-0.2~\textrm{V}$ (blue), $U_{\textrm{g}2}=-0.15~\textrm{V}$ (green), and $U_{\textrm{g}2}=-0.1~\textrm{V}$ (red). The horizontal dashed lines are the heating temperature for channels with infinite width for the gate voltage $U_{\textrm{g}2}$ of corresponding color. The gray region shows the electron transport channel and the vertical dashed lines separate the two stripes with different carrier density from the voltage probes left and right from the channel. }
\end{figure}

Having determined $\tau_\textrm{e}(\epsilon)$ and $\tau_\textrm{i}(\epsilon)$, we calculate the temperature profile, see Fig.~\ref{fig7}, across the sample at the position where the voltage probes are attached. If the energy diffusion lengths $l_i$ were much smaller than the transport-channel widths, then the temperature profile would be a step function (dashed curves in Fig.~\ref{fig7}) with $T_i = T^\textrm{bulk}_i$ (gray area in Fig.~\ref{fig7}). In our sample, however, the channel width is of the same order of magnitude as $l_i$ (for $U_{\textrm{g}i}=-0.2~\textrm{V}/0~\textrm{V}/{+0.2}~\textrm{V}$ the energy diffusion length is $l_i = 16.8~\mu\textrm{m} / 58.2~\mu\textrm{m} / 117.7~\mu\textrm{m}$). As a consequence, we find a strong $y$-dependence of the temperature (solid lines in Fig.~\ref{fig7}), which does not approach the asymptotic bulk values $T^\textrm{bulk}_i$ in the two stripes of the electron channel. Changing the gate voltage $U_{\textrm{g}2}$ in stripe $2$ changes the energy diffusion length $l_2$ there and, thus, $T^\textrm{bulk}_2$. However, it also affects $T^\textrm{bulk}_1$ since for fixed $I_{\textrm{sd}}$ the electric field $E_x$ changes when the carrier density and, thus, the overall resistance is modified.

\begin{figure}
 \includegraphics[width=8cm]{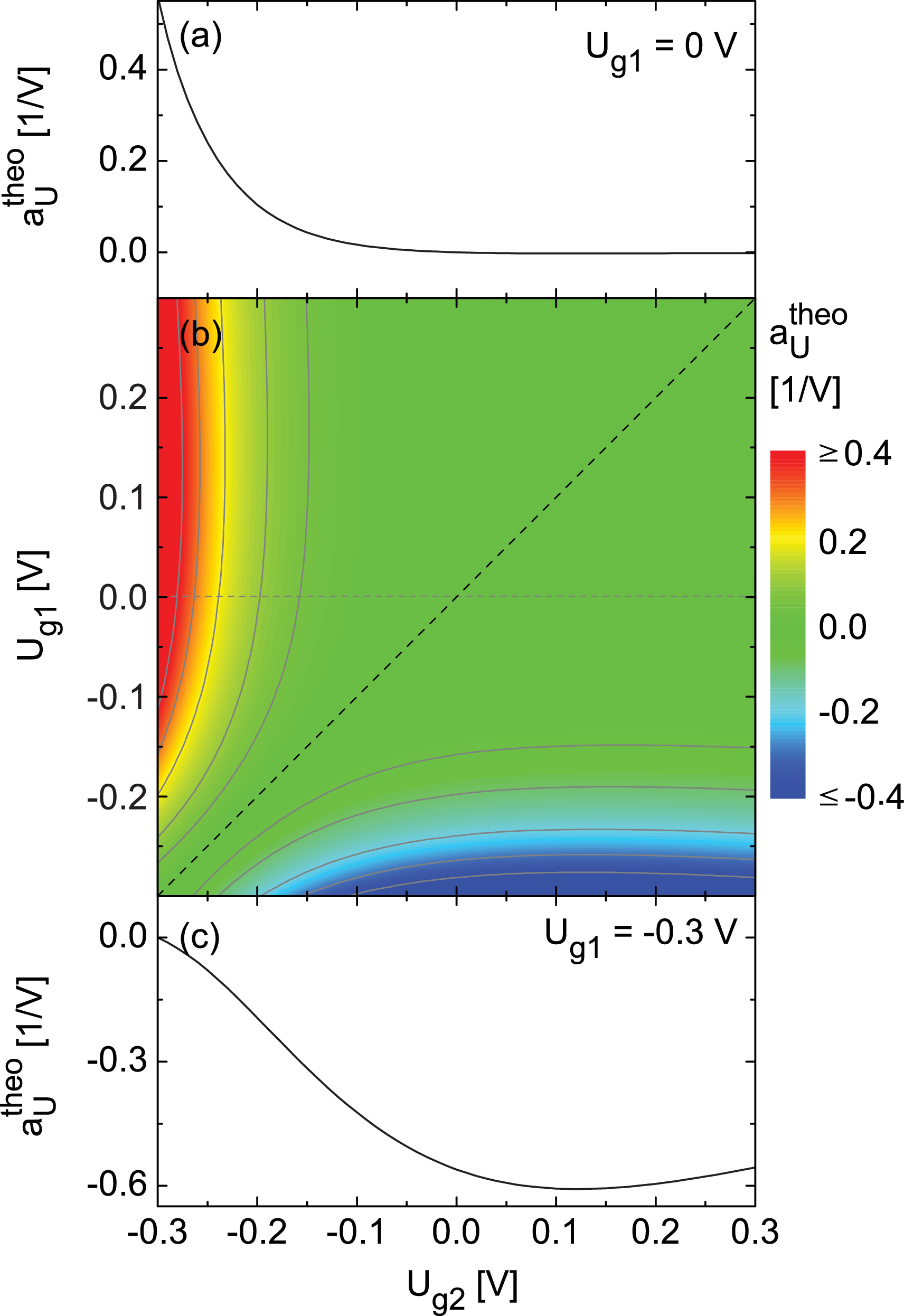}
 \caption{\label{fig8}(a) Calculated rectification efficiency $a_U^{\textrm{theo}}$ as a function of $U_{\textrm{g}2}$ for $U_{\textrm{g}1} = 0~\textrm{V}$. (b) Contour plot of the calculated rectification efficiency $a_U^{\textrm{theo}}$ as a function of $U_{\textrm{g}1}$ and $U_{\textrm{g}2}$. The distance between the contour lines is $0.1~\textrm{V}^{-1}$ with additional contour lines at $\pm 0.05~\textrm{V}^{-1}$. (c) $a_U^{\textrm{theo}}$ in dependence on $U_{\textrm{g}2}$ for $U_{\textrm{g}1} = -0.3~\textrm{V}$. This trace matches the cross section of the contour plot in (b) at $U_{\textrm{g}1} = -0.3~\textrm{V}$.}
\end{figure}

Finally, we plug the temperature profile into Eq.~\eqref{MottInt} to obtain the thermoelectric voltage. To translate the experimentally applied source-drain voltage $U_{\textrm{sd}}$ into the electric field $E_x$ entering the theory, we take an (experimentally estimated) value of $R_0= 450~\Omega$ for the contact resistances from the source and drain contacts. The result of our simulation for the same parameters as in Fig.~\ref{fig3} is shown in Fig.~\ref{fig8}. We find a very good agreement between $a_U^{\textrm{exp}}$ and $a_U^{\textrm{theo}}$ as a function of the gate voltages $U_{\textrm{g}1}$ and $U_{\textrm{g}2}$. In particular, we reproduce the non-monotonic dependence of the rectification efficiency on the density modulation, see Fig.~\ref{fig8}(c). (Note that the exact position of the calculated maximum depends on the contact resistance $R_0$.)

The non-monotonic behavior of $a_U(U_{\textrm{g}1},U_{\textrm{g}2})$ can be explained intuitively within the framework of the diffusion thermopower model by two contrary dependencies. On the one hand, no rectification can be observed for symmetrically biased gates $U_{\textrm{g}2} = U_{\textrm{g}1} = -0.3~\textrm{V}$ so that the efficiency starts from zero (at $U_{\textrm{g}2} = -0.3~\textrm{V}$) and  increases as the gate voltage is increased. On the other hand, the thermoelectric effect decreases with increasing carrier density as the resistivity and, correspondingly, the Joule heating decreases. This eventually leads to a decrease of the overall effect, as observed in the range $U_{\textrm{g}2} > 0.1~\textrm{V}$. 

\begin{figure}
 \includegraphics[width=8cm]{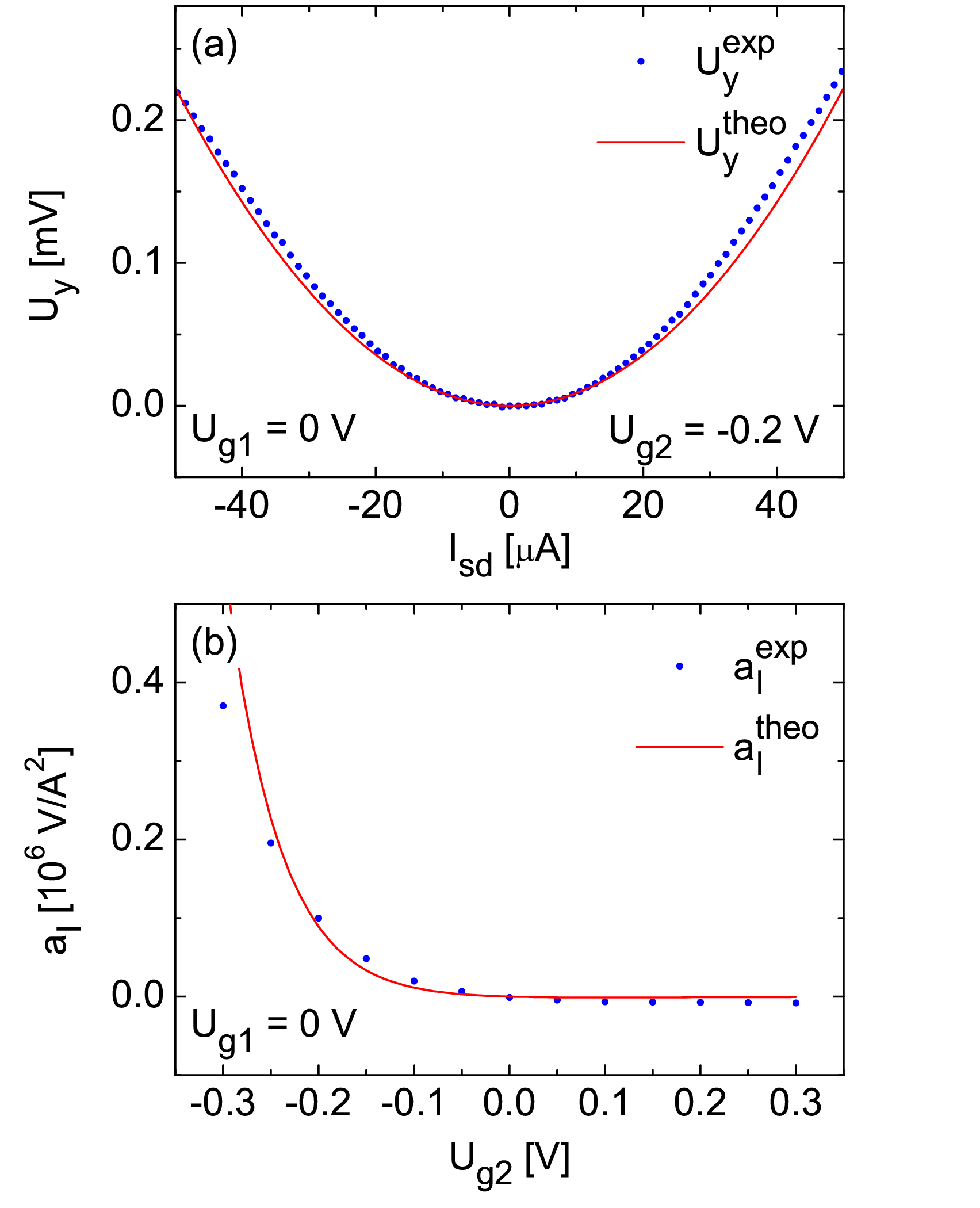}
 \caption{\label{fig9}Comparison between the measured and the calculated rectification effect: (a) Dependence of the measured transverse voltage $U_{y}^{\textrm{exp}}$ (blue dots) and the calculated transverse voltage $U_{y}^{\textrm{theo}}$ (red trace) on the heating current $I_{\textrm{sd}}$ for $U_{\textrm{g}1} = 0~\textrm{V}$ and $U_{\textrm{g}2} = -0.2~\textrm{V}$. (b) Dependence of the measured rectification efficiency $a_U^{\textrm{exp}}$ (blue dots) and the calculated rectification efficiency $a_U^{\textrm{theo}}$ (red trace) on $U_{\textrm{g}2}$.}
\end{figure}

For a quantitative comparison between theory and experiment, we want to avoid the uncertainty associated with the estimate of the contact resistance $R_0$. To this end, we repeat the measurement of the transverse voltage but relate it to the applied current $I_{\textrm{sd}}$ rather than the voltage $U_{\textrm{sd}}$. This defines the rectification efficiency $a_I=U_y/I_{\textrm{sd}}^2$. Figure \ref{fig9} shows a comparison between the experimental and the calculated data. In Fig.~\ref{fig9}(a) the transverse voltage $U_{y}^{\textrm{exp}}$ (blue dots) and the calculated thermoelectric voltage $U_{y}^{\textrm{theo}}$ (red trace) are plotted as a function of $I_{\textrm{sd}}$ for $U_{\textrm{g}1} = 0~\textrm{V}$ and $U_{\textrm{g}2} = -0.2~\textrm{V}$. We find that $U_{y}^{\textrm{theo}}$ shows the same parabolic dependence on $I_{\textrm{sd}}$ as $U_{y}^{\textrm{exp}}$. A comparison of the measured, $a_I^{\textrm{exp}}$, and calculated, $a_I^{\textrm{theo}}$, rectification efficiency as a function of gate voltage is presented in Fig.~\ref{fig9}(b). It should be pointed out that no adjustable parameter have been used in the calculation, since all relevant parameters were independently determined on the homogeneous reference sample B. The small quantitative deviation can be explained by the approximations that were made both in the experimental evaluation and in the theoretical treatment. Also, phonon effects, such as phonon drag, which can play a role in AlGaAs heterostructures,\cite{Fletcher1986, Fletcher1988, Ruf1988, Chickering2009} have been neglected here.

In view of the diffusion thermopower model, it is now also possible to better understand why the rectification efficiency $a_U^{\textrm{exp}}$ and the mobility $\mu_0$ exhibit such different temperature dependencies (see Fig.~\ref{fig4}). Both quantities are affected by $\tau_\textrm{i}$, which has a $1/\textrm{T}$ dependence on the temperature,\cite{Okuyama1989} and $\tau_\textrm{e}$, which is almost constant in the temperature range of interest. The relaxation time $\tau$ that determines the mobility $\mu_0$ is given by Matthiessen’s rule $\tau^{-1}(\epsilon) = \tau^{-1}_\textrm{e}(\epsilon) + \tau^{-1}_\textrm{i}(\epsilon)$, and because $\tau_\textrm{e} \ll \tau_\textrm{i}$, the temperature dependence of $\tau_\textrm{i}$ will only little affect the overall mobility. The energy diffusion length, on the other hand, is proportional to $\sqrt{\tau_\textrm{i} \tau_\textrm{e}}$ (see Eq.~\eqref{energydiffusion}), so that the temperature dependence of $\tau_\textrm{i}$ will affect the rectification efficiency considerably, regardless of how small $\tau_\textrm{e}$ is compared to $\tau_\textrm{i}$. The slight increase of $a_U^{\textrm{exp}}$ at low temperatures may be attributed to an effective increase of $\tau_\textrm{i}$ due to the fact that only scattering processes in which the transferred energy is larger than $k_B T$ contribute to $\tau_\textrm{i}$, i.e, with increasing $T$ less processes contribute.

The experiment and its explanation within the diffusion thermopower model have a number of interesting consequences. Because of the non-monotonic $a_U(U_{\textrm{g}1},U_{\textrm{g}2})$ dependence, the rectification is not ''transitive'', i.e. $a_U(U_{A},U_{B}) + a_U(U_{B},U_{C}) \neq a_U(U_{A},U_{C})$. Therefore, in samples with multiple stripes, the rectification can be cascaded to improve the output voltage. If such a sample is periodic, e.g. with 3 sets of stripes $...ABCABCABC...$, it would constitute a ratchet-type potential (see e.g. Olbrich \textit{et al.}\cite{Olbrich2011} and references therein), where an agitation along the stripes would induce a current/voltage \emph{perpendicular} to the driving signal. Our present setup with a carrier density profile $ABCA$ (voltage probe - first stripe - second stripe - voltage probe, see also \ref{fig1}(b)) constitutes the smallest unit of such a transverse ratchet.

Going back to the introductory remarks, our work also bridges the two concepts for rectification, i.e. change of material vs. non-centrosymmetric guidance of electrons. The two parallel stripes can be considered as two different (but tunable!) materials, which have been interfaced to induce rectification that does not rely on ballistic effects but rather on diffusion thermopower. Nevertheless, we do observe finite size effects. This is a feature that necessarily appears in ballistic rectifiers as well, although with a different characteristic length. While the width of our sample is much larger than the elastic mean free path (the characteristic length for ballistic transport), it is smaller than the energy diffusion length $l$. This leads to a strong reduction of the temperature difference (see Fig.~\ref{fig7}) and, therefore, the rectification efficiency.

In conclusion, we have demonstrated tunable rectification in a density-modulated two-dimensional-system. The polarity and value of the rectified transverse voltage depend on the direction and magnitude of the density modulation of the carrier densities inside the electron channel. We show that the observed data is in good qualitative and quantitative agreement with a diffusion thermopower model. Our findings have a number of consequences for the understanding of so-called ballistic rectifiers and related structures.

{\it Acknowledgements.} -- We acknowledge stimulating discussions with Daniel Urban and financial support from DFG via SPP 1285.

\end{document}